# Fermi level and bands offsets determination in insulating (Ga,Mn)N/GaN structures

L. Janicki[1], G. Kunert[2,3], M. Sawicki[4], E. Piskorska-Hommel[5], K. Gas[5,6], R. Jakiela[4], D. Hommel[2,3], and R. Kudrawiec[1,*]

[1]*Faculty of Fundamental Problems of Technology, Wroclaw University of Science and Technology, Wybrzeże Wyspiańskiego 27, 50-370 Wrocław, Poland*

[2]*Wroclaw Research Center EIT+ Sp. z o.o., ul. Stabłowicka 147, 54-066 Wrocław, Poland*

[3]*Institute of Solid State Physics, University of Bremen, Otto-Hahn-Allee 1, 28359 Bremen, Germany*

[4]*Institute of Physics Polish Academy of Sciences, Al. Lotnikow 32-46, 02-668 Warsaw, Poland*

[5]*Institute of Low Temperature and Structure Research, Polish Academy of Sciences, Institute W. Trzebiatowski, ul. Okolna 2, 54- 422 Wroclaw, Poland*

[6]*Institute of Experimental Physics, University of Wroclaw, pl. Maxa Borna 9, 50-204 Wroclaw, Poland*

*corresponding author: robert.kudrawiec@pwr.edu.pl

The Fermi level position in (Ga,Mn)N has been determined from the period-analysis of GaN-related Franz-Keldysh oscillation obtained by contactless electroreflectance in a series of carefully prepared by molecular beam epitaxy GaN/Ga$_{1-x}$Mn$_x$N/GaN(template) bilayers of various Mn concentration $x$. It is shown that the Fermi level in (Ga,Mn)N is strongly pinned in the middle of the band gap and the thickness of the depletion layer is negligibly small. For $x > 0.1\%$ the Fermi level is located about 1.25 - 1.55 eV above the valence band, that is very close to, but visibly below the Mn-related Mn$^{2+}$/Mn$^{3+}$ impurity band. The accumulated data allows us to estimate the Mn-related band offsets at the (Ga,Mn)N/GaN interface. It is found that most of the band gap change in (Ga,Mn)N takes place in the valence band on the absolute scale and amounts to -0.028±0.008 eV/% Mn. The strong Fermi level pinning in the middle of the band gap, no carrier conductivity within the Mn-related impurity band, and a good homogeneity enable a novel functionality of (Ga,Mn)N as a semi-insulating buffer layers for applications in GaN-based heterostuctures.

Dilute ferromagnetic semiconductors are in the focus of research interest since they combine functionalities of semiconductors and magnetic materials providing a prolific playground for both basic research and technology viable applications[1,2]. Among many considered systems, a ferromagnetic guise of GaN – (Ga,Mn)N in which manganese substitutes randomly gallium – would constitute a major technological advance due to the already dominating role of group III nitrides in photonics and high power electronics. An intensive research, seeded by the seminal paper by Dietl et al.[3] has led, however, to somehow contradicting results concerning the possible nature of magnetism of (Ga,Mn)N, pointing decisively to the specifics of Mn distribution within GaN host lattice and/or to the presence of donor-like centers or defects as to the main factors determining the properties of the material[4]. The most promising reports of ferromagnetism persisting to well above the room temperature in (Ga,Mn)N[5–8] have now lost their initial impact since the lack of any spintronic functionality reported to date. It is now accepted that isolated ferromagnetic (FM) mesoscopic volumes with a very high Mn concentration, which develop in otherwise very diluted paramagnetic environment, give rise to the overall FM-like appearance. On the other hand, in samples which were grown according to a very carefully prepared growth protocol resulting in random distribution of Mn cations and a relatively small concentration of donor-like centers, a low temperature uniform FM has been documented[9–12]. The experimentally established dependence of the Curie temperature ($T_C$) on Mn concentration, $T_C \sim x^{2.2}$, supports the short-range FM $Mn^{3+}$ - $Mn^{3+}$ superexchange scenario[13]. The same exponent describes the dependence of spin-glass freezing temperatures on composition in Mn- and Co-doped dilute magnetic semiconductors in which the antiferromagnetic superexchange is the established spin coupling mechanism[14]. However, to make the model plausible and in an accordance with other experimental findings[15,16], it has been postulated that in (Ga,Mn)N the Fermi level is pinned by the mid-gap $Mn^{2+}/Mn^{3+}$ impurity band, and that the highly insulating character results from a strong $p – d$ coupling driven localization of Mn-derived impurity-band holes[17] which prevails over the long range ordering, at least to currently available $x \leq 13\%$. This scenario has been convincingly supported by tight-binding and Monte-Carlo computations[11,12]. Our approach enables to independently test earlier reports of the position of the Fermi level in (Ga,Mn)N and, furthermore, allows to determine the band offset between GaN and (Ga,)MnN. Both are important parameters for an integration of the material in high power nitride devices.

In this paper we propose the application of modulation spectroscopy to accurately establish the Fermi level position in insulating $Ga_{1-x}Mn_xN$ layers with $0.1 < x < 10\%$ exhibiting low temperature FM properties with $T_C < 12$ K. The method, as detailed later, relies on the analysis of the period of Franz-Keldysh oscillation present in optical spectra measured for thin GaN cap layers overgrown on the researched $Ga_{1-x}Mn_xN$ films. The optical studies allowed us to establish a strong mid-gap Fermi level pinning and a lack of a depletion layer in (Ga,Mn)N. Those results, in combination with already established insulating character[18,19] and a sizable dielectric strength of 5 MV/cm[20], strongly favor (Ga,Mn)N as an insulating buffer material for applications in (high power) nitride devices. At the same time, the accumulated results allow us to estimate the Mn-related band offsets in (Ga,Mn)N which we find to predominantly take place in the valence band amounting to $28 \pm 8$ meV per %Mn.

**RESULTS AND DISCUSSION**

Contactless electroreflectance (CER), which is one of electromodulation spectroscopy techniques, is an absorption-like technique allowing precise studies of direct optical transitions in semiconductor materials[21,22] and low dimensional heterostructures[23]. It also provides a way to detemine the built-in electric field in semiconductor structures *via* an analysis of Franz-Kieldysh oscillation (FKO) which appears in the CER spectrum for samples with medium or strong electric fields[24–28]. Studies of such a built-in electric field in specially designed semiconductor structures allow to determine the Fermi level position at the relevant semiconductor surface (i.e., the surface potential)[26,27,29], which is an indispensable information for design of structures in which the electric field distribution needs to be tailored for a desired device operation, e.g. in transistor structures[28].

In this study CER spectroscopy is applied to determine the Fermi level position in single phase $Ga_{1-x}Mn_xN$ with $x$ ranging from 0.1 to 7%. All the measurements were performed at room temperature in air ambient. For the reasons detailed below the (Ga,Mn)N layers have been capped with undoped GaN layers. In the first set of structures the thickness $d$ of the GaN caps is kept constant at $d = 60$ nm, in the second one $d$ varies from 31 to 170 nm. The intended layout of the investigated structures is presented in the inset to Fig. 1 which shows a typical x-ray diffraction pattern and the relevant structural simulation indicating a very good epitaxial composition of our structures. Details on the growth and on the results of the structural and chemical characterization of the samples are given in Methods. As a reference, to support this study, single (Ga,Mn)N layers already investigated in Refs. 10 and 12 were used.

We start the analysis of our results by considering the reference monolayers. Figure 2 shows CER spectra measured at room temperature for GaN template (panel a) and the reference single (Ga,Mn)N layers (panel b). For GaN template the band gap-related transition is observed at ~3.42 eV and the Fabry-Perot (F-P)-related CER signal is visible below 3.4 eV[28,30]. The Fabry-Perot signal, marked as F-P, is observed because of a large contrast of refractive indices at the following interfaces air/(Ga,Mn)N and GaN(template)/sapphire. The observation of a band gap-related signal in the CER spectrum means that a surface band bending (a surface electric field) exists near the sample's surface and this bending is modulated in CER measurements. Generally, it is the presence of a surface electric field that is responsible for the appearance of the CER signal. However, for (Ga,Mn)N layers the intensity of the band gap-related transition decreases with $x$ and gets totally quenched for $x > 4\%$ or perhaps even earlier since the CER signal in samples with $x \leq 4\%$ looks more like a F-P signal than a band-to-band one. Thus, the absence of the band-gap-related transition in these layers indicates that there is almost no depletion at the surface. This means that in high-$x$ case the Fermi level must be pinned at a bulk-related characteristic energy $\gamma$ both in the bulk of the layer and at the surface.

To determine the position of $\gamma$ an artificial depletion layer is introduced to (Ga,Mn)N by capping it with a thin undoped GaN overlayer in which a surface-like electric field ($F$) can build up. The required potential difference across the GaN cap is provided by the difference between the electrochemical potentials at the GaN/(Ga,Mn)N interface (set by the position of $\gamma$) and at the GaN free surface ($\phi$, set by the position of surface states) as it is sketched in Fig. 3 (a). The former is assumed to be equal to the Fermi level position in (Ga,Mn)N since the surface depletion layer for uncapped high-$x$ (Ga,Mn)N layers is negligibly thin according to CER measurements performed on the reference layers. The latter is equally important. We can assume that the Fermi level at the surface of the whole structure is at the same position as for a single GaN layer. Calculations show that for a perfect $c$-plane Ga-polar GaN surface with a 2x2 Ga$_{T4}$ reconstruction, two singularities exist in the surface density of states[31]. The low energy one is located ~1.7 eV below the conduction band and is fully occupied by electrons for a perfect crystal. The high energy one starts about 0.6 eV below the conduction band and continues upward. Whereas the latter singularity pins the Fermi level for $n$-type GaN, for a semi-insulating and $p$-type GaN the surface Fermi level position should be set by the low energy singularity. This has been observed *in situ* by XPS on $n$-type and $p$-type GaN[32]. Exposure to air can change the surface reconstruction and adsorption of oxygen to GaN surface can occur[33]. However, the surface Fermi level of air exposed GaN(0001) shows similar properties to bare 2x2 Ga$_{T4}$ surface with two distinct singularities of surface density of states. The earlier performed CER studies on GaN/GaN:Si and GaN/GaN:Mg structures yielded $\phi = 0.3 - 0.6$ eV[26,27,34] and $\phi = 1.7$ eV[34] for $n$- and $p$-type structures, respectively. Because Mn acts as a deep acceptor in GaN the value $\phi = 1.7$ eV is adopted in this study. Finally, the strength of $F$ in GaN cap is set by its thickness $d$. Let us underline here that the placement of the undoped GaN cap not only provides a depletion-like layer to (Ga,Mn)N but also assures a precise knowledge of its width. Therefore, the Fermi level position in (Ga,Mn)N can be readily obtained from the condition:

$$\gamma = E_g - Fd - \phi \qquad (1)$$

where $E_g = 3.42$ eV is the GaN band gap.

The expression describing the shape of the CER spectra in presence of the built-in electric field is given below[35]:

$$\frac{\Delta R}{R} \propto \exp\left[\frac{-2\Gamma\sqrt{E-E_g}}{(\hbar\theta)^{3/2}}\right] \cdot \cos\left[\frac{4}{3}\left(\frac{E-E_g}{\hbar\theta}\right)^{3/2} + \varphi\right] \cdot \frac{1}{E^2(E-E_g)},$$

$$(\hbar\theta)^3 = \frac{e^2\hbar^2 F^2}{2\mu}, \qquad (2)$$

and from the cosine part the extrema of Franz-Keldysh oscillation (FKO) are given by:

$$n\pi = \phi + \frac{4}{3}\left[\frac{(E_n - E_g)}{\hbar\theta}\right]^{\frac{3}{2}}. \qquad (3)$$

In the above equations $\hbar\theta$ is the electro-optic energy, $\Gamma$ is the linewidth, $\varphi$ is an arbitrary phase factor, $\mu$ is the electron-hole reduced mass (assumed to be 0.2 $m_0$[36]), $n$ is the index of the $n$-th extremum, and $E_n$ is the corresponding energy. A plot of $(E_n-E_g)^{3/2}$ versus $n$ yields then a straight line which slope is proportional to $F$.

Room temperature CER spectra measured for the two sets of GaN/(Ga,Mn)N structures with a constant and varying thickness of the GaN cap are shown in Figs. 4 and 5, respectively. Clearly, as intended, a band-to-band transition followed by FKO is observed for all samples. For the structure with a thick GaN cap (Fig. 5 (f)) the built-in electric field is weak and, therefore, an excitonic contribution is strong in this case, see the dashed box. However, even in this case, the contribution of a band-to-band transition followed by a FKO is visible, see CER spectrum multiplied by a factor of 2 and shifted vertically. Although the period of the FKO is showing sizable variations across the whole range of structures, a clear pattern emerges for samples with $x > 0.1\%$. As indicated in Fig. 4 (b-e) the period is nearly constant for structures having constant width of the GaN cap and it shows a decreasing trend with increasing $d$, as shown in Fig. 5. The presence of FKO in CER spectra confirms the sensitivity of the applied method to built-in electric fields. The strength of fields is established from the plots of the FKO extrema positions versus their index, presented in Fig. 6 for both sets of structures. For each sample a clear linear dependency is seen and the values of $F$ extracted from the relevant slopes (Eq. 2) are shown in the legends.

The established values of $F$ present in GaN/(Ga,Mn)N structures with $x > 0.1\%$ allow us to determine the Fermi level position from Eq. 1. Figure 7 (a) summarizes our findings presenting the established dependence of the Fermi level energy in (Ga,Mn)N on $x$. The presented data do not exhibit any specific variation with respect to the GaN cap width meaning that the Fermi level pinning on the GaN/(Ga,Mn)N interface is predominantly determined by the bulk of the (Ga,Mn)N layer and, therefore, strongly connected with the Mn level position in the host lattice. Our findings indicate that the Fermi level in (Ga,Mn)N is located ~1.25 - 1.55 eV above the valence band edge, i.e. in a close proximity to the Mn-related band which is located between 1.4 to 1.8 eV above the valence band edge in (Ga,Mn)N[2,15,37–39]. This conclusion is consistent with recent studies of the valency of Mn atoms in (Ga,Mn)N films which was found to be 2.4 eV[40] as well as with ellipsometric assessment of the Fermi level position in (Ga,Mn)N and corresponding theoretical calculations[41]. Such conditions of the Fermi level position in (Ga, Mn)N are certainly very unfavorable for achieving the carrier mediated ferromagnetism in this material, at room temperature in particular, and they advocate strongly for the superexchange $Mn^{3+}$ - $Mn^{3+}$ coupling postulated in Refs. 10 and 12 to explain rather low temperature ferromagnetism in these samples.

The Fermi level position is markedly different for low-$x$ structures, exemplified here by the GaN(60nm)/$Ga_{0.999}Mn_{0.001}$N structure. As illustrated in Fig. 4 (a) the FKO is much weaker and of an opposite phase than the oscillation observed for other samples. This indicates that the bands are bent in a opposite direction[42], i.e., the surface electric field in the GaN cap layer has an opposite sign than in the high-$x$ structures. In addition, the CER signal shape shown in Fig. 4 (a) indicates that the character of the transition can be more excitonic than the band-to-band one, hence, rendering the determination of $F$ in this structure less accurate. Assuming, however, a band-to-band character of this transition we have obtained that the corresponding strength of the built-in electric field in low-$x$ GaN/(Ga,Mn)N samples does not exceed 18 kV/cm. This means that the Fermi level at the interface between GaN and low-$x$ (Ga,Mn)N is located closer to the conduction band edge than in high-$x$ structures and its position is similar to the Fermi level position at the GaN surface. Taking into account that nominally undoped GaN is $n$-type due to native defects (i.e., the Fermi level is near the conduction band) an increase in the Mn incorporation into the GaN host should shift the Fermi level from the conduction band towards the Mn impurity band and, therefore, for Mn concentration < 0.1 % the Fermi level should be located somewhere between the conduction band and the Mn impurity band.

So far we have neglected the Mn-related change in the band gap of (Ga,Mn)N alloy in our analysis. This approach has led to a somewhat puzzling result that for a higher $x$ the Fermi level is located at least a thousand Kelvin below the Mn level and, as Fig. 7 (a) clearly indicates, it gets consistently deeper with increasing $x$. This inconsistency can be reconciled by recalling that the (Ga,Mn)N band gap should be a function of $x$ for alloy-like concentrations of Mn. Indeed, as postulated in Ref. 17, there is experimental evidence that the (Ga,Mn)N band gap increases with $x$ at a rate of 0.027 eV per % of Mn[43]. Therefore, by fixing the (Ga,Mn)N Fermi level at a constant position above the valence band we can translate the variation of $E_F$ with $x$ (Fig. 7(a)) into a down-shift of the (Ga,Mn)N valence band with respect to the parent GaN one and, therefore, determine the valence band offset, $\Delta_{VB}$, between GaN and (Ga,Mn)N. Such a case is sketched in Fig. 3 (b) where $x$-dependent positions of both conduction and valence band edges are marked as dotted lines. Here, for each $x$, the $\Delta_{VB}$ is readily obtained from the adequately modified condition (1): $\gamma + \Delta_{VB} = E_g - Fd - \phi$, where now $\gamma$ is fixed at 1.5 eV above the valence band independently of $x$, and $\Delta_{CB}$, the conduction band offset, is obtained accordingly from: $\Delta_{VB} + \Delta_{CB} = 0.027x$ eV/% Mn. As illustrated in Fig. 7 (b), under such assumptions most of the band gap change in (Ga,Mn)N takes place in the valence band on the absolute scale and amounts to $\Delta_{VB}$ = -0.028±0.008 eV/% Mn which is the first estimation of this quantity given so far.

Considering functionalities of the (Ga,Mn)N alloy it is worth noting that (Ga,Mn)N layers studied in this paper are highly resistive like samples studied in Refs. [18] and [19]. The high resistivity of (Ga,Mn)N samples and the negligibly small surface depletion layer mean that this material can serve as an excellent insulator in nitride devices. Such insulators are particularly desirable for planar AlGaN/GaN field effect transistors where it is of a paramount importance to isolate the conductivity of the two-dimensional electron gas from that of the substrate or/and a buffer layer. Currently, the semi-insulating character of GaN layers is achieved either by maintaining a low residual doping or by an intentional doping by deep acceptors like C or Fe[44–46]. Those are, however, of a limited use since Fe atoms tend to segregate into Fe-rich nanocrystals which leads to deterioration of their carrier trapping capabilities[47] whereas the incorporation of C leads to formation of dislocations and numerous self-compensating defects[48].

**CONCLUSIONS**

In conclusion, a reliable method to study the Fermi level position in dilute ferromagnetic semiconductors such as (Ga,Mn)N has been presented. The method is based on the determination of the strength of the built-in electric field in an insulating cap layer overgrown on the studied material by means of an analysis of the period of Franz-Keldysh oscillation observed in contactless electroreflectance spectroscopy. Using this approach, it has been shown that the Fermi level in (Ga,Mn)N with $x > 0.1$% is located about $1.25 - 1.55$ eV above the valence band edge, i.e. close to the Mn-related band which, according to previous studies[2,15,37–39], is located $\sim 1.4 - 1.8$ eV above the valence band edge in GaN. Simultaneously, for the first time for this compound, the data allowed an estimation of the Mn-related band offsets between GaN and (Ga,Mn)N pointing out that the majority of the band gap change in (Ga,Mn)N takes place in the valence band amounting to $-0.028 \pm 0.008$ eV/% Mn. The strong mid-gap pinning of the Fermi level established in this study combined with good solubility of Mn in GaN, the lack of a surface/interface depletion layer, and strong insulating properties univocally predestinate (Ga,Mn)N to take over Fe or C doped GaN as a highly resistive buffer in GaN-based transistors and other applications (including high power) where highly resistive materials with the lattice constant very close to GaN are desirable.

**METHODS**

**Sample growth.** The GaN/(Ga,Mn)N structures studied here were grown by molecular beam epitaxy (MBE) in a VEECO EPI930 MBE chamber equipped with a radio-frequency plasma source. Single-side polished [0001]-oriented (*c*-plane) sapphire substrates with an about 2 μm thick GaN layer deposited by metal-organic vapor phase epitaxy were used as templates. Prior to the growth, the backside of the substrates was covered by a 1 μm thick Ti layer to improve the temperature homogeneity during growth. The substrates were cleaned in an ultrasonic bath and subsequently degassed in a vacuum system at temperatures of up to 800 °C. Two sets of samples were grown at temperatures between $730 - 760$ °C following the details outlined in Ref. [10]. In the first set, a ~200 nm thick (Ga,Mn)N layer of various Mn concentrations ($x$ = 0.1, 0.35, 1.4, 1.6, and 6.5%) was capped by a $d$ = 60 nm thick undoped GaN layer. In the second set the thickness of the GaN cap was varied from 31 to 170 nm. The deposition of both (Ga,Mn)N and GaN layers was monitored *in situ* by reflection high energy electron diffraction (RHEED). All samples showed bright, sharp, and streaky reflections in RHEED during the whole growth time, indicating a smooth layer growth. Single (Ga,Mn)N layers were used as a reference for CER measurements, to support this study. These samples were already investigated in Refs. [10] and [12].

**Structural studies.** The Mn profile and its concentration in the structures was checked by secondary ion mass spectrometry and Mn content was independently confirmed by superconducting quantum interference device magnetometry[49]. In addition, X-ray diffraction measurements were performed on all samples to evaluate their quality, layer thickness and confirm their Mn content and profile independently. A Philips X'Pert MRD high resolution X-ray diffractometer with a monochromator and an Eulerian cradle was used to perform structural analyses of the samples. The Mn concentrations were determined from Vegard's law by simulations of Omega-2Theta scans of the (0002)-reflex of GaN and (Ga,Mn)N. The lattice parameter of (Ga,Mn)N was taken from Ref. [10].

**Contactless electroreflectance.** For CER measurements the samples were mounted in a capacitor with a half-transparent top electrode made from a copper-wire mesh[42]. (Ga,Mn)N samples were fixed to the bottom copper electrode by silver paste. The distance between the sample surface and the top electrode was ~0.5 mm. A single grating 0.55 meter focal-length monochromator and a photomultiplier were used to disperse and detect the light reflected from the samples. Phase-sensitive detection of the CER signal was performed using a lock-in amplifier. Other relevant details on CER technique and measurements can be found in Ref. [42].

## AUTHOR INFORMATION

**Corresponding Author**

*E-mail: robert.kudrawiec@pwr.edu.pl

**Author Contributions**

L. J. performed contactless electroreflectance measurements, G. K. fabricated the samples, performed XRD measurements and analysis, E. P.-H. did structural studies and analysis, K. G. performed SQUID measurements and analysis, R. J. performed SIMS measurements and analysis. M.S. and D. H. analyzed the experimental data. R. K. analyzed the experimental data, supervised the project, and together with M.S. wrote the manuscript. All the authors discussed the results and reviewed the manuscript.

**Notes**

The authors declare no competing financial interest.

## ACKONOWLEDGMENTS

This study has been supported by the National Science Centre (Poland) through OPUS Grants No. 2011/03/B/ST3/02633 and 2013/09/B/ST3/04175, FUGA Grant No. 2014/12/S/ST3/00549, by the EU 7th Framework Programmes: CAPACITIES project REGPOT-CT-2013-316014 (EagLE) and by the Wroclaw Research Centre EIT+ within the project "The Application of Nanotechnology in Advanced Materials" - NanoMat (P2IG.01.01.02-02-002/08) co-financed by the European Regional Development Fund (operational Programme Innovative Economy 1.1.2).

**Figure Captions**

Figure 1. Exemplary measurement and simulation of the 2Theta/Omega peak relative to the GaN (0002) peak of one sample. Inset: Intended sample structure. Deviations were determined by XRD.

Figure 2. Room temperature contactless electroreflectance spectra of a GaN template (a) and (Ga,Mn)N layers of various Mn concentrations (b).

Figure 3. A schematic representation of the bands alignments in GaN/(Ga,Mn)N (hetero)structures (a) assuming that the GaN band gap, $E_g$, does not change in (Ga,Mn)N, (b) allowing an increase of the band gap in (Ga,Mn)N. $\Delta_{VB}$ and $\Delta_{VC}$ are valence and conduction band discontinuities, respectively. $\gamma$ is the Fermi level ($E_F$) distance from the valence band in (Ga,Mn)N, $\phi$ is the position of $E_F$ pinning surface states in relation to the conduction band. The field F is set by the GaN cap width $d$ and the energy difference $\Delta E$.

Figure 4. Room temperature contactless electroreflectance spectra of GaN(60nm)/(Ga,Mn)N structures of various Mn concentrations. The observed CER signal at ~3.42 eV followed by FKO is associated with the band-to-band absorption in the GaN cap. Natural numbers indicate extrema of FKO. For samples with low Mn concentration the Fabry-Perot signal, marked as F-P, is observed below 3.4 eV.

Figure 5. Room temperature contactless electroreflectance spectra of GaN(d)/(Ga,Mn)N structures of various thicknesses of GaN cap layer and various Mn concentration. The observed CER signal at ~3.42 eV followed by FKO is associated with the band-to-band absorption in the GaN cap. Natural numbers indicate extrema of FKO.

Figure 6. Analysis of GaN-related Franz-Keldysh oscillation for GaN(60nm)/(Ga,Mn)N structures of various Mn concentrations (a) and GaN(d)/(Ga,Mn)N structures of various thicknesses of GaN cap layer and various Mn concentrations (b). Electric fields obtained from this analysis are given in the legend.

Figure 7. (a) Position of the Fermi level in (Ga,Mn)N determined for GaN(60nm)/(Ga,Mn)N structures of various Mn concentration (open points) and GaN(d)/(Ga,Mn)N structures of various thicknesses of GaN cap layer and various Mn concentration (full points) with the assumption of no band gap change in (Ga,Mn)N upon the incorporation of Mn. The grey bar indicates the position of Mn-related $Mn^{2+}/Mn^{3+}$ impurity band taken from Refs.[2, 15, 34-36]. (b) Analysis of the valence band position in (Ga,Mn)N under the assumption that the Fermi level (Ga,Mn)N is fixed (1.5 eV above the valence band) and that the band gap opening rate in (Ga,Mn)N amounts to 0.027 eV per % Mn [44]. The inaccuracy of the Fermi level position determination in (Ga,Mn)N layers corresponds to sizes of the open and solid squares.

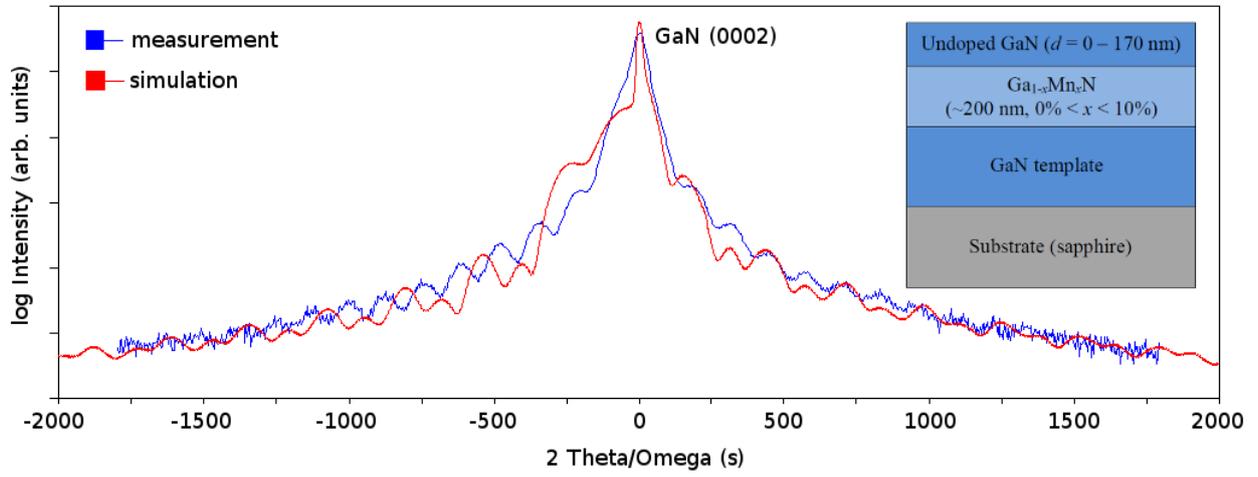

Figure 1. Exemplatory measurement and simulation of the 2Theta/Omega peak relative to the GaN (0002) peak of one sample. Inset: Intended sample structure. Deviations were determined by XRD.

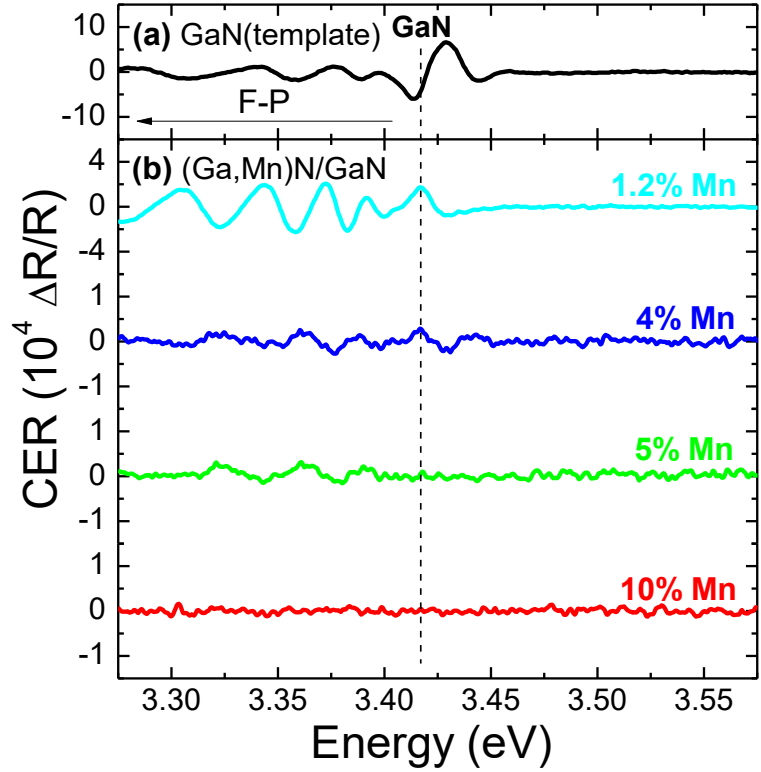

Figure 2. Room temperature contactless electroreflectance spectra of GaN template (a) and (Ga,Mn)N layers of various Mn concentrations (b). The Fabry-Perot signal, marked as F-P, is observed because of a large contrast of refractive index at the following interfaces air/(Ga,Mn)N and GaN(template)/sapphire.

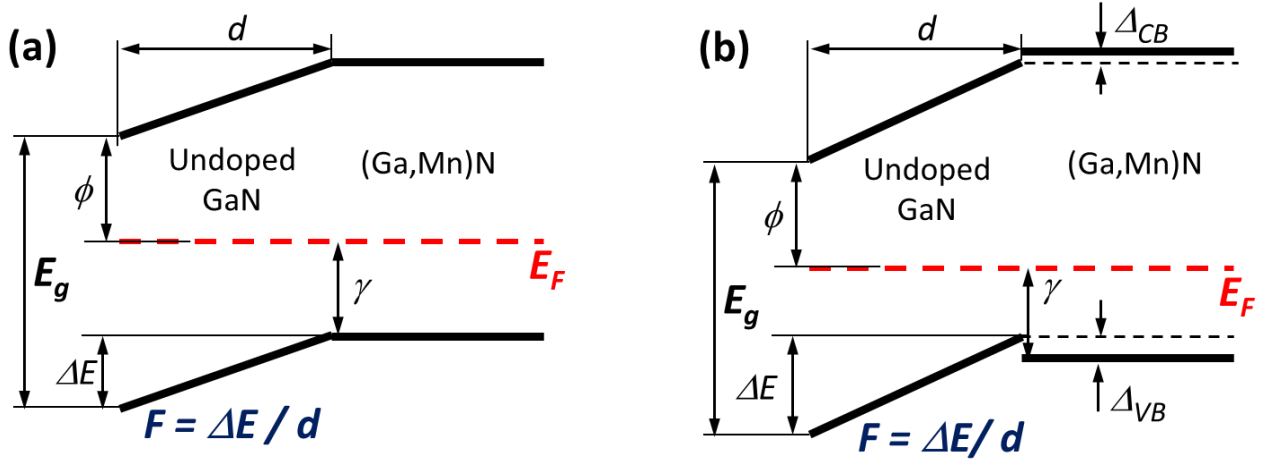

Figure 3. A schematic representation of the bands alignments in GaN/(Ga,Mn)N (hetero)structures. (a) assuming that the GaN band gap, $E_g$, does not change in (Ga,Mn)N, (b) allowing an increase of the band gap in (Ga,Mn)N. Here bands discontinuities (off-sets) appear at the GaN/(Ga,Mn)/N interface ($\Delta_{VB}$ and $\Delta_{VC}$ for the valence and conductivity bands, respectively). In both cases it is assumed that the Fermi level ($E_F$) is pinned at a certain energy $\gamma$ above the valence band edge in (Ga,Mn)N and at the surface density of states positioned at energy $\phi$ below the conduction band edge at the free surface of GaN . The strength of the developing electric field $F$ in the undoped GaN layer is set by its width $d$ and the energy difference $\Delta E$ set by the different Fermi level positions at the GaN free surface and at GaN/(Ga,Mn)N interface.

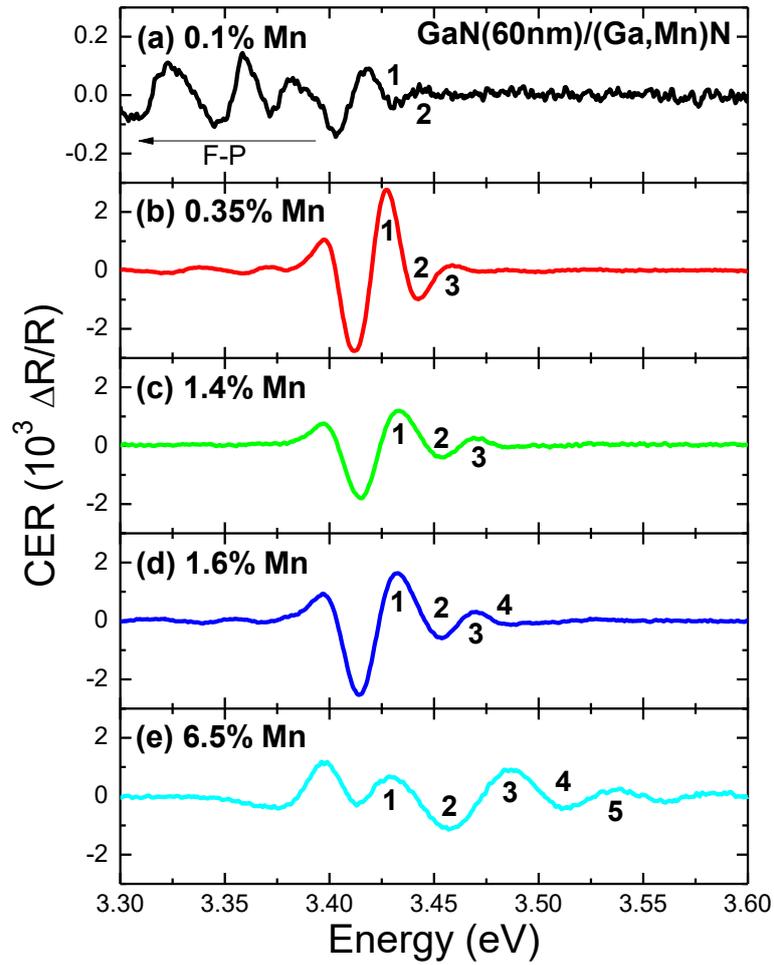

Figure 4. Room temperature contactless electroreflectance spectra of GaN(60nm)/(Ga,Mn)N structures of various Mn concentrations. The observed CER signal at ~3.42 eV followed by FKO is associated with the band-to-band absorption in the GaN cap. Natural numbers indicate extrema of FKO. For samples with low Mn concentration the Fabry-Perot signal, marked as F-P, is observed below 3.4 eV.
24

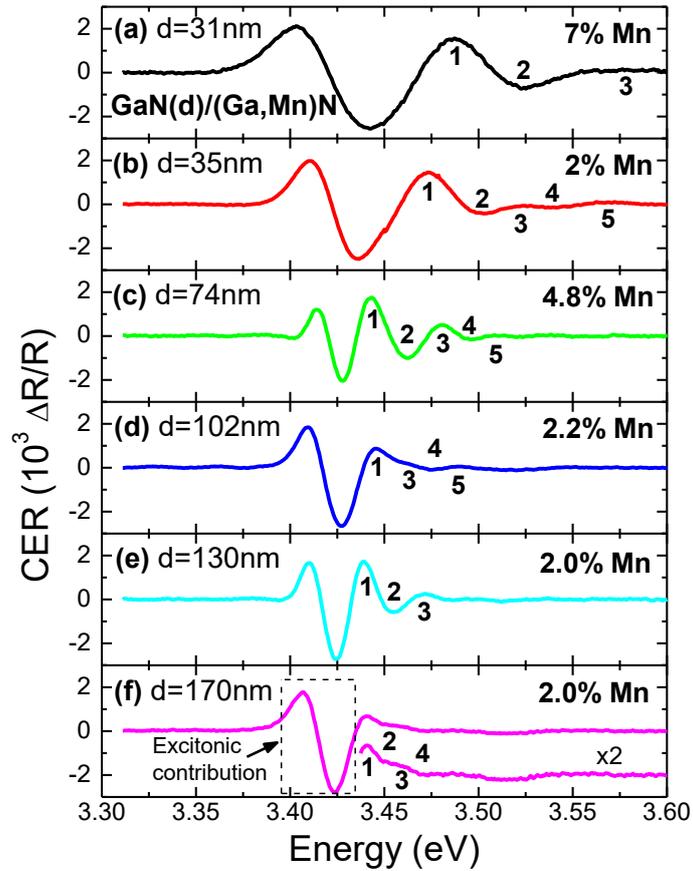

Figure 5. Room temperature contactless electroreflectance spectra of GaN(d)/(Ga,Mn)N structures of various thicknesses of GaN cap layer and various Mn concentration. The observed CER signal at ~3.42 eV followed by FKO is associated with the band-to-band absorption in the GaN cap. Natural numbers indicate extrema of FKO. For the structure with a thick GaN cap (panel f) the built-in electric field is weak and, therefore, an excitonic contribution is strong in this case, see the dashed box. However, even in this case, the contribution of a band-to-band transition followed by a FKO is visible, see CER spectrum multiplied by a factor of 2 and shifted vertically.

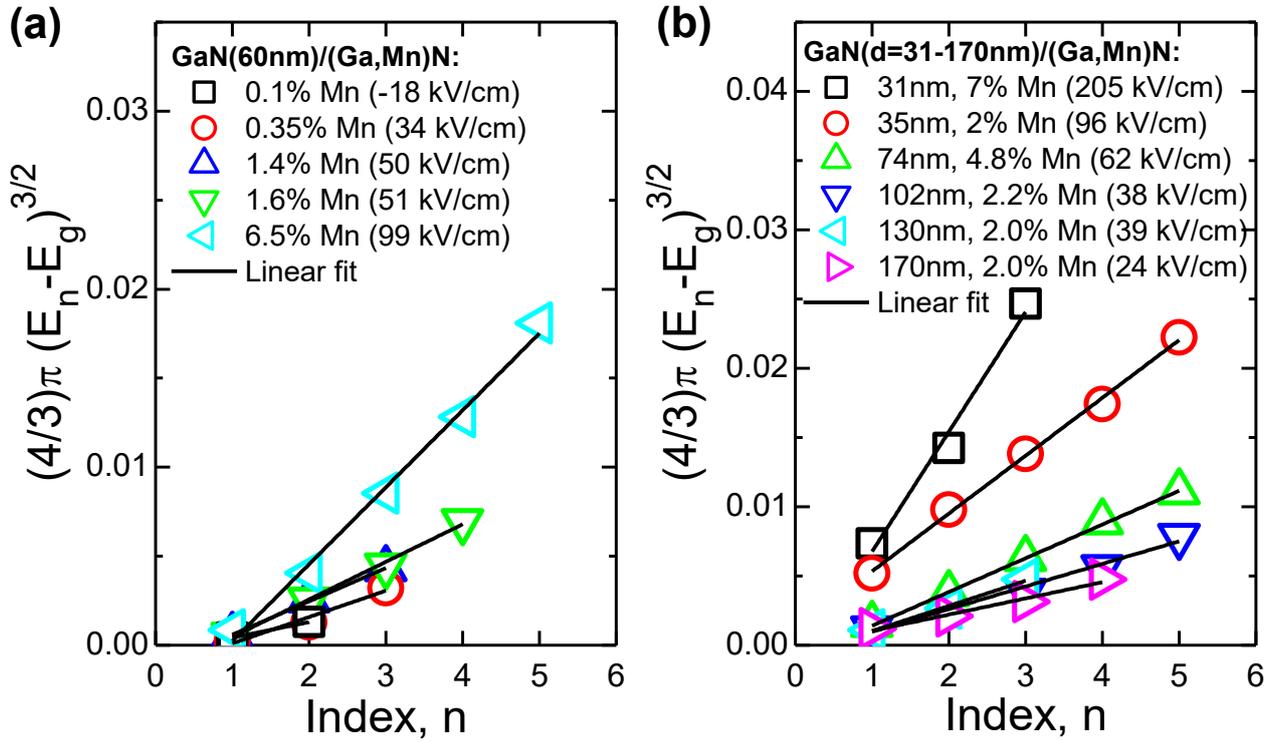

Figure 6. Analysis of GaN-related Franz-Keldysh oscillation for GaN(60nm)/(Ga,Mn)N structures of various Mn concentrations (a) and GaN(d)/(Ga,Mn)N structures of various thicknesses of GaN cap layer and various Mn concentrations (b). Electric fields obtained from this analysis are given in the legend.

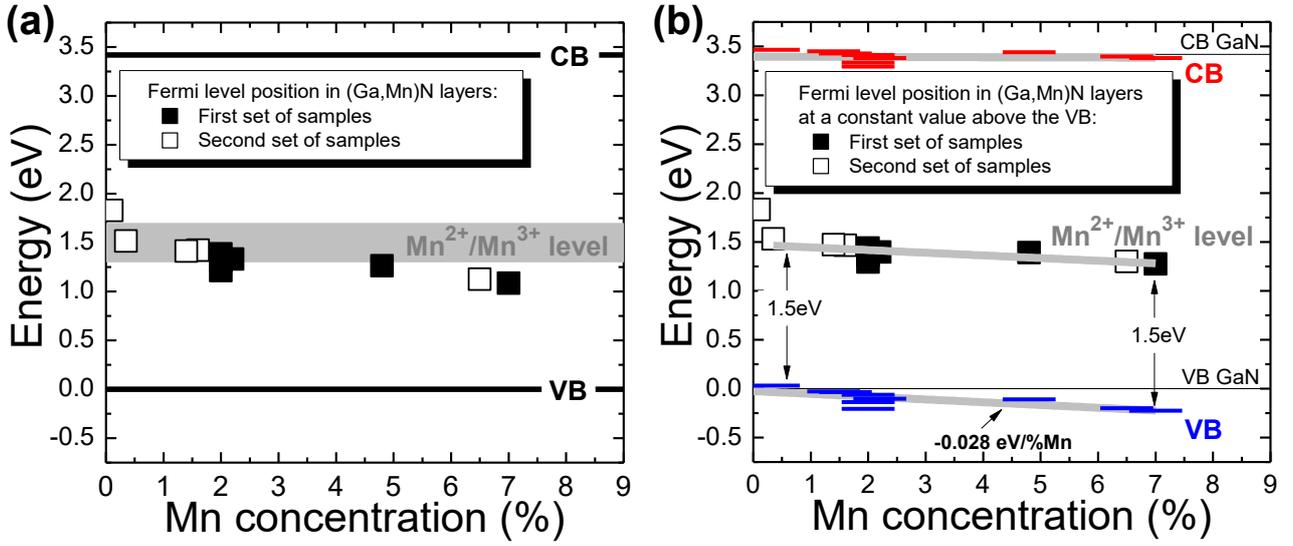

Figure 7. (a) Position of the Fermi level in (Ga,Mn)N determined for GaN(60nm)/(Ga,Mn)N structures of various Mn concentration (open points) and GaN(d)/(Ga,Mn)N structures of various thicknesses of GaN cap layer and various Mn concentration (full points) with the assumption of no band gap change in (Ga,Mn)N upon the incorporation of Mn. The horizontal thick grey bar indicates the position of Mn-related $Mn^{2+}/Mn^{3+}$ impurity band taken from Refs.[2, 15, 34-36]. (b) Analysis of the valence band position in (Ga,Mn)N under the assumption that the Fermi level (Ga,Mn)N is fixed (1.5 eV above the valence band) and that the band gap opening rate in (Ga,Mn)N amounts to 0.027 eV per % Mn [44]. The determined valence band shift is -0.028±0.008 eV/% Mn. The conduction band shift is negligibly small and equals -0.001±0.008 eV/% Mn. The inaccuracy of determination of the Fermi level position in (Ga,Mn)N layers corresponds to sizes of the open and solid squares. The given above valence and conduction band position errors are not indicated explicitly in panel (b) for clarity, they are of the same order as the inaccuracy of determination of the Fermi level position.